\documentclass[paper=a4]{scrartcl}

\usepackage{xr}
\usepackage{bm}
\RequirePackage{amsthm,amsmath,amsfonts,amssymb}
\usepackage{natbib}
\usepackage{booktabs}
\usepackage{xcolor}
\usepackage{graphicx}
\usepackage{hyperref}
\usepackage{tikz}
\usetikzlibrary{arrows.meta,positioning,fit,decorations.markings}
\usepackage{mathtools}
\usepackage{enumitem}

\usepackage{scrlayer-scrpage}
\ihead{Bayesian spatial+}
\ohead{\pagemark}
\usepackage{authblk}

\title{\LARGE Bayesian spatial+: A joint model perspective}
\date{\vspace{-1em}}

\author[1]{Isa Marques\thanks{Corresponding author.}}
\author[1]{Paul F.V. Wiemann}

\affil[1]{\normalsize Department of Statistics, The Ohio State University, Columbus, OH, USA}

\newcommand{\red}[1]{\textcolor{red}{#1}}

\mathtoolsset{showonlyrefs}

\theoremstyle{plain}

\theoremstyle{definition}

\theoremstyle{remark}

\begin{document}
	\maketitle
	\begin{abstract}
		\small
		Spatial confounding is a common issue in spatial regression models, occurring when spatially varying covariates correlate with the spatial effect included in the model.
This dependence, particularly at high spatial frequencies, can introduce bias in regression coefficient estimates when combined with smoothing penalties.
 The spatial+ framework is a widely used two-stage frequentist approach that mitigates spatial confounding by explicitly modeling and removing the spatial structure in the confounding covariate, then using the corresponding residuals in the second-stage model for the response.
However, it does not propagate first-stage uncertainty, does not discuss a general inferential framework, and, crucially, cannot guarantee that covariate residuals and spatial effects in the response model are free of shared high-frequency structure, so confounding may persist.
We propose Bayesian spatial+, a joint modeling approach that simultaneously addresses these limitations.
Our framework naturally propagates uncertainty and enables straightforward posterior inference, while ensuring separation of spatial frequencies through specialized joint priors on smoothness parameters.
We further introduce a cut-feedback strategy that prevents feedback between model components from reintroducing confounding.
Simulation studies and real-world applications show substantial gains in bias reduction and interval coverage relative to existing approaches.
Notably, in our comparisons, Bayesian spatial+ is the only method for which credible interval coverage remains stable as the sample size increases.

	\end{abstract}
	\section{Introduction}
\label{sec:intro}

Spatial regression models are commonly employed to account for residual spatial dependence in a regression model after considering observed covariates. This residual spatial dependence can be thought of as the result of unobserved, spatially varying covariates. When these unobserved covariates are correlated with the observed covariates, the resulting regression coefficients may be biased, and their uncertainty underestimated.

A standard approach to adjust for unobserved spatial covariates is to approximate their collective effect using an unknown function $f$, defined over the spatial domain. Consider observations $y(\bm{s}_i)$ within the spatial domain $\mathcal{S}$ at locations $\bm{s}_i \in \mathcal{S} \subseteq \mathbb{R}^2$, for $i = 1, \ldots, n$. These observations can be modeled as
\begin{equation}\label{eq:motivation}
	y(\bm{s}_i) = x(\bm{s}_i) \beta + f(\bm{s}_i) + \varepsilon(\bm{s}_i), 
\end{equation} 
where $x(\bm{s}_i)$ is a spatially-varying covariate, $\beta$ is the associated coefficient, and each $\varepsilon(\bm{s}_i)$ is i.i.d. normally distributed  with mean zero and constant variance.  Throughout, we assume that covariates and response are centered at zero, such that intercepts can be omitted.

Unfortunately, spatial dependence between observed and unobserved covariates is often reflected in the spatial dependence between $f$ and $x$. This correlation may lead to biased estimates of regression coefficients and inflated standard errors.
 This phenomenon, first identified by \citet{clayton1993spatial} and later termed ``spatial confounding'' by \citet{reich2006effects}, has been extensively studied, with numerous articles addressing its causes and mitigation strategies \citep[e.g.][]{reich2006effects, paciorek2010importance, thaden2018structural, khan2020restricted, dupont2022spatial+, marques2022mitigating, guan2022spectral, dupont2023demystifying}.

Recent advances have identified two key mechanisms driving spatial confounding: smoothing and spatial frequencies. Spatial confounding vanishes when no smoothing is applied to the spatial effect \citep{dupont2022spatial+, dupont2023demystifying}, however smoothing is desirable for predictive purposes. When smoothing is present, the degree of bias depends critically on the type of spatial frequencies shared between the covariates and the spatial effect.

Spatial frequencies refer to the rate of variation of spatial patterns across the domain: higher spatial frequencies correspond to rapidly varying patterns, while lower spatial frequencies correspond to slowly varying, smooth patterns. Shared high frequencies between spatial effect and covariate are more problematic because spatial smoothing penalties disproportionately penalize high-frequency components in the spatial effect, leading to greater bias in the covariate's estimated coefficient. In contrast, shared low frequencies are less influential as they experience weaker penalization \citep{dupont2023demystifying}. These findings align with earlier insights by \citet{paciorek2010importance}, who emphasized the risks of spatial effects operating at ``smaller scales'' than spatial covariates, and the role of smoothing as a bias inducer.
Recently, in the spectral domain, \citet{guan2022spectral} developed a model that assumes that spatial confounding vanishes at high spatial frequencies. 
A variety of work has followed similar ideas \citep[e.g.][]{makinen2022spatial, urdangarin2023one}.
Note that broadly speaking, spatial scale and spatial frequency are inversely related: larger spatial scales correspond to lower spatial frequencies, and vice versa. In this paper, we use the term spatial frequency to align with recent work by \citet{guan2022spectral} and \citet{dupont2023demystifying}, and because it seems more precise.

In response to these challenges, numerous methods have been proposed to mitigate spatial confounding. Restricted spatial regression (RSR) \citep{reich2006effects, hanks2015restricted} was an early remedy, imposing orthogonality between the spatial effect and the covariates, thus leading to the same mean estimates for $\beta$ as a non-spatial (NS) model. 
However, if the response is spatially varying, the NS model is misspecified and the estimates of $\beta$ might be biased.
In addition, recent work \citep{khan2020restricted, zimmerman2021deconfounding} shows that, contrary to expectations, the posterior variance of $\beta$ can be smaller for RSR than for NS, ultimately also leading to worse coverage rates and higher Type-S errors \citep{gelman2000type}.

Following these findings, alternative approaches have emerged, such as those by \citet{dupont2022spatial+}, \citet{marques2022mitigating}, or \citet{guan2022spectral}.
A particularly influential method is the spatial+ framework \citep{dupont2022spatial+}, which has inspired numerous extensions \citep{marques2022spatial+, reich2022spatial+, urdangarin2022evaluating, urdangarin2023one, schmidt2022discussion}. Spatial+ uses a two-stage frequentist approach: a spatial model is first fitted to the covariate, and the residuals from this model replace the covariate in a second-stage regression. 
Thus, ideally, the residuals that replace the covariate in the second stage equation  should have sufficient \textit{unconfounded} high spatial frequencies or non-spatial information such that $\beta$ can still be identified.

While the spatial+ framework represents a significant advance in addressing spatial confounding, its frequentist implementation faces important limitations.  This has led to subsequent modifications  \citep{urdangarin2023one, dupont2023demystifying}.
For instance, spatial+ cannot guarantee there is no leakage of confounded spatial frequencies from the first stage to the second stage model. To improve this, \citet{dupont2023demystifying} employ the same basis functions for spatial effects in both modeling stages but cap the number of high frequencies and set the first-stage smoothness parameter to zero -- though the choice of this cap remains non-trivial. Moreover, inference procedures for spatial+ have not yet been formally developed. \citet{schmidt2022discussion} propose a Bayesian variant using Gaussian processes instead of splines, which facilitates inference but does not resolve the issues related to frequency leakage. Finally, \citet{marques2022spatial+} suggest a joint Bayesian formulation to propagate uncertainty from the first stage to the second.

This paper extends the spatial+ framework into a Bayesian setting through a joint model.
By introducing a novel joint prior for smoothness-related components, we address key limitations of the frequentist approach.
Our method offers several advantages: 
(i) enhanced uncertainty quantification (UQ) through joint estimation that properly propagates uncertainty across model stages;
(ii) straightforward inference for all model parameters within the Bayesian paradigm; and
(iii) prior restrictions that guarantee no frequency leakage -- rather than restrictions on the posterior, like methods such as RSR.
Specifically, our Bayesian framework employs a joint prior that controls the smoothness-related parameters of spatial effects for both the covariate and response.
This eliminates the need for users to specify a frequency cap -- a challenging task.
Additionally, it mitigates issues associated with highly rank-deficient spatial effects, which can inflate uncertainty in $\beta$ and diminish predictive performance.
Importantly, we introduce a novel approach to Bayesian feedback that prevents oversmoothing of the second-stage spatial effect, thereby avoiding bias, leading to much improved results.

The paper is organized as follows.
Section~\ref{sec:sp2} reviews the spatial+ method of \citet{dupont2022spatial+}.
Section~\ref{sec:bsp2} introduces the Bayesian spatial+ model, presenting our main contributions:  novel joint priors for smoothness parameters that eliminate the need for user-specified frequency caps and  prevent frequency leakage and a novel cutting feedback mechanism designed for this particular model.
The extension to multiple covariates is also discussed.%
Section~\ref{sec:ss} evaluates the model's performance in two simulation studies, demonstrating substantial improvements in bias reduction and coverage over existing methods.
Finally, Section~\ref{sec:application} presents applications to meteorological and forestry data that validate our approach in real-world settings, before we conclude in Section~\ref{sec:conclusion}.

\section{Spatial+ in a nutshell}
\label{sec:sp2}

Consider a single observed covariate $x$ and assume that both covariate and response are centered, so that we can omit intercepts. Our starting point is the spatial+ model of \citet{dupont2022spatial+}, which is formulated as a two-stage regression
\begin{subequations}
\begin{align}
  x(\bm{s}_i) &=  f^x(\bm{s}_i) + \varepsilon^x(\bm{s}_i), \label{eq:spatial-plus-x}\\
  y(\bm{s}_i) &= \beta \bigl( x(\bm{s}_i) -  \hat{f}^x(\bm{s}_i) \bigr) + f(\bm{s}_i) + \varepsilon(\bm{s}_i)
               = \beta \hat{\varepsilon}^x(\bm{s}_i) + f(\bm{s}_i) + \varepsilon(\bm{s}_i),   \label{eq:spatial-plus-y}
\end{align}
\end{subequations}
where $f$ and $f^x$ are smooth functions on the spatial domain $\mathcal{S} \subseteq \mathbb{R}^2$ evaluated at locations $\bm{s}_i \in \mathcal{S}$, and $\varepsilon^x(\bm{s}_i) \stackrel{iid}{\sim} N(0,\sigma_x^2)$, $\varepsilon(\bm{s}_i) \stackrel{iid}{\sim} N(0,\sigma^2)$ are independent random errors.
The term $\hat{f}^x(\bm{s})$ denotes the fitted values of $f^x(\bm{s})$ from \eqref{eq:spatial-plus-x}, and $\hat{\varepsilon}^x(\bm{s}_i) = x(\bm{s}_i) - \hat{f}^x(\bm{s}_i)$ are the corresponding residuals.
In \citet{dupont2022spatial+}, and subsequent work, $f$ and $f^x$ are estimated using thin plate regression splines \citep[TPRS,][]{wood2003thin}.

The key idea is that the residuals $\hat{\varepsilon}^x$ are dominated by high-frequency spatial and non-spatial variation in $x$ and therefore serve as a proxy for the unconfounded component of the covariate.
Using these residuals in the second stage is reasonable as long as they retain enough information to identify $\beta$.
Thus, the performance of spatial+ hinges on two conditions. Condition~1 (C1) $\hat{\varepsilon}^x$ must contain sufficient non-spatial or high-frequency information to identify $\beta$ in the second stage, and Condition~2 (C2) the high-frequency components in $\hat{\varepsilon}^x$ must not be confounded with the spatial effect $f$.

C1 may fail for very smooth spatial covariates with little non-spatial variation, such as climate variables like air temperature, where $\hat{\varepsilon}^x$ may not be informative enough to identify $\beta$. More critically, C2 cannot be guaranteed without additional assumptions as the original spatial+ framework does not prevent leakage of shared high-frequency components between the covariate residuals and the spatial effect in the response \citep[see][]{dupont2023demystifying, urdangarin2023one}.

\section{Bayesian Spatial+}
\label{sec:bsp2}

In this section, we develop a Bayesian joint modeling version of spatial+, designed to
\begin{itemize}
	\itemsep0em 
	\item[(i)] propagate uncertainty from the covariate model to the response model,  
	\item[(ii)] enforce separation of high spatial frequencies between covariate residuals and the response level spatial effect through a joint smoothness prior, while controlling feedback in the joint model, and
	\item[(iii)] provide straightforward posterior inference for all model components.
\end{itemize}
We first introduce the joint model specification, then describe the basis representation and reparameterization we use.
Building on this, we construct a joint smoothness prior that guarantees frequency separation between the residual and the response-level spatial effect.
We then discuss feedback in the joint model and strategies to control it, before we turn to estimation and extensions.

\subsection{A joint model specification}
We consider a response $y(\cdot)$ defined on a spatial domain $\mathcal{S} \subseteq \mathbb{R}^2$ and, initially, a single spatial covariate $x(\cdot)$ (we lift this restriction in Section~\ref{sec:bsp-extensions}).
As before, we assume that covariate and response are centered, so that intercepts can be omitted. Let
 $$
  \bm{y} = \bigl(y(\bm{s}_1),\dots,y(\bm{s}_n)\bigr)^\top
  \qquad\text{and}\qquad
  \bm{x} = \bigl(x(\bm{s}_1),\dots,x(\bm{s}_n)\bigr)^\top.
$$
The Bayesian spatial+ model replaces the plug-in residuals from the frequentist
two stage approach by a joint model for both the covariate and response. We model
\begin{subequations}
\begin{align}
  x(\bm{s}_i) &= f^x(\bm{s}_i) + \varepsilon^x(\bm{s}_i), 
  & \varepsilon^x(\bm{s}_i) &\stackrel{iid}{\sim} N(0,\sigma_x^2), \label{eq:bayes_spatialplus_x}\\
  y(\bm{s}_i) &= \beta \varepsilon^x(\bm{s}_i) + f(\bm{s}_i) + \varepsilon(\bm{s}_i),
  & \varepsilon(\bm{s}_i) &\stackrel{iid}{\sim} N(0,\sigma^2), \label{eq:bayes_spatialplus_y}
\end{align}
\end{subequations}
for $i=1,\dots,n$, where $f^x$ and $f$ are smooth spatial effects and $\varepsilon(\bm{s}_i)$ and $\varepsilon^x(\bm{s}_i)$ are the latent residuals from the response model and covariate model, respectively. Importantly, $\varepsilon^x(\bm{s}_i)$ is not sampled; it is obtained as the difference  $\varepsilon^x(\bm{s}_i) = x(\bm{s}_i) - f^x(\bm{s}_i)$, since treating noisy draws of $\varepsilon^x(\bm{s}_i)$ as covariates would inflate uncertainty and can bias regression coefficients \citep{carroll2006measurement}.

In vector form, with $\bm{\varepsilon}^x = (\varepsilon^x(\bm{s}_1),\dots,\varepsilon^x(\bm{s}_n))^\top$ and
$\bm{\varepsilon} = (\varepsilon(\bm{s}_1),\dots,\varepsilon(\bm{s}_n))^\top$, this can be written as
\begin{align*}
  \bm{x} &= \bm{f}^x + \bm{\varepsilon}^x
  &\text{and}
  &&\bm{y} &= \beta \bm{\varepsilon}^x + \bm{f} + \bm{\varepsilon},
\end{align*}
where $\bm{f}^x = \bigl(f^x(\bm{s}_1),\dots,f^x(\bm{s}_n)\bigr)^\top$ and
$\bm{f} = \bigl(f(\bm{s}_1),\dots,f(\bm{s}_n)\bigr)^\top$.

This joint formulation makes explicit that the same latent residual process $\bm{\varepsilon}^x$ drives both the covariate model \eqref{eq:bayes_spatialplus_x} and the covariate contribution to the response model \eqref{eq:bayes_spatialplus_y}.
Compared to the frequentist spatial+ model, there is no plug-in estimate $\hat{f}^x$ and no deterministic residual vector.
Instead, uncertainty about $f^x$ -- and hence about the residual process $\varepsilon^x$ -- is fully propagated to inference on $\beta$.

We complete the model by assigning priors to the regression coefficient, variance parameters and spatial effects. Specifically, we let
$$
  \beta \sim N(0, \theta_\beta^2), \qquad
  \sigma \sim \mathrm{HN}(0,\theta_\sigma^2), \qquad \sigma_x \sim \mathrm{HN}(0, \theta_{\sigma_x}^2),
$$
where $\mathrm{HN}(0,\theta_*^2)$ denotes a half-normal distribution with scale
$\theta_* > 0$.
The spatial effects $f$ and $f^x$ are represented via a common spline basis, with corresponding coefficients and smoothness parameters.
The basis representation and the associated reparameterization are described in the next subsection, while the joint smoothness prior is developed in Section~\ref{sec:joint-prior}.

\subsection{Basis representation and reparameterization}
\label{sec:basis}

\subsubsection{Basis representation}
A popular way of approximating spatially varying functions is via bivariate basis function approaches \citep{ramsay2002spline, wood2003thin, sangalli2013spatial, ugarte2017one}.
We represent $f$ and $f^x$ in \eqref{eq:bayes_spatialplus_x} and \eqref{eq:bayes_spatialplus_y} using a common system of spatial basis functions.
Let $\{B_j(\cdot)\}_{j=1}^d$ denote such a basis on $\mathcal{S}$ and define
$$
  f(\bm{s}) = \sum_{j=1}^d \gamma_j B_j(\bm{s}), 
  \qquad 
  f^x(\bm{s}) = \sum_{j=1}^d \gamma^x_j B_j(\bm{s}),
$$
with coefficient vectors $\bm{\gamma} = (\gamma_1,\dots,\gamma_d)^\top$ and $\bm{\gamma}^x = (\gamma^x_1,\dots,\gamma^x_d)^\top$.
Let $\bm{Z}$ be the $n\times d$ design matrix with entries $Z_{ij} = B_j(\bm{s}_i)$, so that $\bm{f} = \bm{Z}\bm{\gamma}$ and $\bm{f}^x = \bm{Z}\bm{\gamma}^x$.

Following standard penalized spline practice, we associate a quadratic penalty matrix $\bm{K}$ with the basis $\{B_j\}$ and use a smoothness parameter $\lambda$ to control the overall roughness of $f$.
In frequentist penalized regression, this yields a penalty term $\lambda \bm{\gamma}^\top \bm{K} \bm{\gamma}$.
In the Bayesian setting, the corresponding (potentially improper) prior is
\begin{equation}
\bm{\gamma} \mid \tau^2 \propto (\tau^2)^{-r/2} 
  	\exp \left(-\frac{1}{2\tau^2} \bm{\gamma}^\top \bm{K} \bm{\gamma}\right), \label{eq:prior-gamma1}
\end{equation}
with $r = \mathrm{rank}(\bm{K}) \leq d$ and $\lambda = \sigma^2/\tau^2$ linking the frequentist smoothing parameter to the prior variance $\tau^2$ \citep[see][]{nychka2000spatial}.
An analogous construction is used for $\bm{\gamma}^x$ with variance $\tau_x^2$ and smoothing parameter $\lambda_x = \sigma_x^2/\tau_x^2$.

\subsubsection{Reparameterization and spatial frequencies}

For many spline bases such as TPRS, the penalty matrix $\bm{K}$ measures the roughness of the function and can be viewed as a discrete approximation to a continuous roughness penalty.
The matrix is typically rank deficient $(r < d)$ as it is setup not to regulate specific low-frequency components such as the linear part.

Therefore, we can split the coefficient vector into penalized and unpenalized components by reparameterizing
\begin{equation}
  \bm{\gamma} = \tilde{\bm{V}} \bm{\gamma}_{\mathrm{pen}} + \tilde{\bm{U}} \bm{\gamma}_{\mathrm{unpen}},
  \label{eq:reparam_gamma}
\end{equation}
where the $d \times r$ matrix $\tilde{\bm{V}}$ and the $d \times (d - r)$ matrix $\tilde{\bm{U}}$ can be obtained form the spectral decomposition of $\bm{K}$. $\tilde{\bm{U}}$ spans the null-space of $\bm{K}$ and the matrices satisfy 
$\tilde{\bm{V}}^\top \bm{K} \tilde{\bm{V}} = \bm{I}_r$ and $\tilde{\bm{U}}^\top \bm{K} \tilde{\bm{U}} = \bm{0}$ \citep[e.g.][]{wand2000comparison, fahrmeir2004penalized}.
By construction, the vector $\bm{\gamma}_{\mathrm{unpen}}$ operates on the unpenalized null space of $\bm{K}$ and while $\bm{\gamma}_{\mathrm{pen}}$ operates on the penalized complement.
We obtain $\bm{\gamma}^\top\bm{K}\bm{\gamma} = \bm{\gamma}_\text{pen}^\top\bm{\gamma}_\text{pen}$ and assign priors
$$
  \bm{\gamma}_{\mathrm{pen}} \mid \tau^2 \sim N(\bm{0}, \tau^2 \bm{I}_r),
  \qquad
  p(\bm{\gamma}_{\mathrm{unpen}}) \propto 1,
$$
yielding a proper prior on the penalized component and a flat prior on the unpenalized part.
The penalized component now has a well-behaved prior, while the flat prior on the unpenalized component remains improper but is generally unproblematic.
Under mild conditions on the rank of the design matrices and the priors on the variance components, the posterior remains proper even if  $\bm{K}$ is rank deficient \citep[see][]{fahrmeir2009propriety, klein2015bayesian}.

The same construction is used for the covariate effect, with
$$
   \bm{\gamma}^x = \tilde{\bm{V}} \bm{\gamma}^x_{\mathrm{pen}} + \tilde{\bm{U}} \bm{\gamma}^x_{\mathrm{unpen}},
$$
and analogous priors for $\bm{\gamma}^x_{\mathrm{pen}}$ and $\bm{\gamma}^x_{\mathrm{unpen}}$ but conditional on $\tau^2_x$ instead of $\tau^2$.
In terms of the implied basis for the spatially varying functions, we obtain
$$
  \bm{f} = \bm{V} \bm{\gamma}_{\mathrm{pen}} + \bm{U} \bm{\gamma}_{\mathrm{unpen}}
  \qquad \text{and} \qquad
  \bm{f}^x = \bm{V} \bm{\gamma}^x_{\mathrm{pen}} + \bm{U} \bm{\gamma}^x_{\mathrm{unpen}},
$$
with $\bm{U} = \bm{Z}\tilde{\bm{U}}$ and $\bm{V} = \bm{Z}\tilde{\bm{V}}$. Although both $\bm{f}$ and $\bm{f}^x$ use the same basis functions, the associated coefficients are identified; see Proposition~S.1 in Supplement~\ref{app:identification_basis} for a proof. Using the same basis functions guarantees that $\bm{f}$ and $\bm{f}^x$  are constructed from the same set of spatial frequencies, which is crucial for comparing and controlling their contributions.

\subsection{A joint smoothness prior and frequency separation}
\label{sec:joint-prior}

\subsubsection{Shrinkage of high frequencies}

To gain insight into the shrinkage, consider the spectral decomposition $\bm{K} = \bm{\Gamma}\bm{\Omega}\bm{\Gamma}^\top$, where the diagonal matrix $\bm{\Omega} = \operatorname{diag}(\omega_1, \dots, \omega_d)$ contains eigenvalues ordered by decreasing magnitude and $\bm{\Gamma}$ contains the corresponding eigenvectors.
Let $\bm{\Omega}^+$ be the diagonal matrix constructed from the first $r$ non-zero eigenvalues and let $\bm{\Gamma}^+$ be the corresponding submatrix of eigenvectors.
Then $\tilde{\bm{V}} = \bm{\Gamma}^+ (\bm{\Omega}^+)^{-1/2}$ while $\tilde{\bm{U}}$ is a basis of the null space of $\bm{K}$ \citep[see][for details]{fahrmeir2004penalized,klein2016scale}.
Consider, an alternative but equivalent parameterization to \eqref{eq:reparam_gamma}.
Writing $\tilde{\bm{\gamma}}_{\mathrm{pen}} = (\bm{\Omega}^+)^{-1/2}\bm{\gamma}_{\mathrm{pen}}$, the prior becomes $\tilde{\bm{\gamma}}_{\mathrm{pen}} \mid \tau^2 \sim N(\bm{0}, \tau^2 (\bm{\Omega}^+)^{-1})$, making explicit that components corresponding to larger eigenvalues $\omega_j$ receive smaller prior variance $\tau^2/\omega_j$.

Each eigenvector $\bm{\delta}_j$ of $\bm{K}$ induces a spatial mode $\bm{m}_j = \bm{Z}\bm{\delta}_j$ and we can decompose the penalized part of $\bm{f}$  as a sum of weighted modes $\bm{Z}\bm{\Gamma}^+\tilde{\gamma}_\text{pen} = \sum_{j=1}^r \tilde{\gamma}_{\text{pen},j} \bm{m}_j$.
For roughness-based penalties, modes associated with larger $\omega_j$ exhibit higher spatial frequency than those with smaller eigenvalues.
Consequently, higher spatial frequency patterns receive strong shrinkage \citep{dupont2023demystifying}.

\subsubsection{A joint prior for frequency separation}

The key idea of Bayesian spatial+ is to use a joint prior on the smoothness-related parameters to enforce that $f^x$ to capture all lower spatial frequencies also present in $f$.
Consequently, the residuals $\varepsilon^x = x - f^x$ contain high-frequency variation that $f^x$ is too smooth to absorb and, thus, the even smoother $f$ cannot absorb, and therefore, providing an unconfounded source of information for identifying $\beta$.

We parameterize smoothness in terms of the usual ratios
$$
  \lambda = \frac{\sigma^2}{\tau^2},
  \qquad
  \lambda_x = \frac{\sigma_x^2}{\tau_x^2}.
$$
Our goal is to impose $\lambda_x < \lambda$, i.e., to ensure that the response effect is smoothed at least as strongly as the covariate effect.
Rather than constraining $\lambda$ and $\lambda_x$ directly, we construct a prior on the variance parameters $(\tau^2, \tau_x^2)$ that
implies this ordering.
Specifically, we set
\begin{equation}
  \tau_x^2 
  = \tau^2 \frac{\sigma_x^2}{\sigma^2} (1+\xi_x^2),
  \qquad
  \xi_x \sim \mathrm{HN}(0, \theta_x^2),
  \label{eq:taux-param2}
\end{equation}
with fixed scale $\theta_x > 0$ in the half-normal prior, so $\lambda_x = \lambda / (1 + \xi_x^2) < \lambda$ almost surely.
Under this prior, $f$ is always at least as smooth as $f^x$, and typically more so.
Consequently, the residuals $\varepsilon^x$ preserve higher-frequency components that
cannot be captured by $f$.

To understand the relevance of this ordering, consider three cases
\begin{enumerate}
	\item If $\lambda < \lambda_x$, then $f^x$ is smoother than $f$, and
  $\varepsilon^x$ may retain spatial frequencies that remain confounded with $f$.

	\item If $\lambda = \lambda_x$, the two spatial effects have identical smoothness.
		In principle this can avoid confounding, but joint estimation tends to be unstable when the data-generating smoothness of $f$ and $f^x$ differ.

	\item If $\lambda > \lambda_x$, the residuals $\varepsilon^x$ retain high-variation that $f$ cannot absorb, reducing spatial confounding. Extreme choices such as $\lambda_x=0$ (no smoothing for $f^x$) guarantee orthogonality at the cost of discarding potentially informative structure in $x$ required for successful identification of $\beta$ (see condition C1 in Section~\ref{sec:sp2}).
\end{enumerate}

Our prior \eqref{eq:taux-param2} implements the third case in a flexible way.
Instead of capping the frequencies in the basis representation, it preserves the full basis for both effects while regulating their relative smoothness through the latent variable $\xi_x$.
Section~\ref{sec:bsp-estimation2} discusses the weakly informative priors placed on ($\sigma$, $\sigma_x$,  $\tau$).
Under mild conditions, all hyperparameters are identified; see Proposition~S.2 in the Supplement for details.

\subsection{Feedback and how to control it}
\label{sec:feedback2}

The joint smoothness prior introduces a deliberate dependence between the covariate and response models.
This dependence is crucial for frequency separation, but it also creates a feedback mechanism in the joint posterior.
Information from the covariate model can influence the effective amount of smoothing in the response model.
In particular, if the this mechanism makes the response model overly smooth then bias will be introduced \citep{paciorek2010importance, dupont2022spatial+, dupont2023demystifying}.

The joint prior enforces $\lambda_x = \lambda/(1+\xi_x^2) < \lambda$ and thus links the covariate-side smoothing to the response-side smoothing via the relationship among $\sigma$, $\tau$ and $\tau_x$.
When the data for $x$ favor very smooth effects (e.g., large $\lambda_x$; see Scenario~2 in Section~\ref{sec:ss1}), the posterior for $(\sigma, \tau)$ may move towards configurations that increase $\lambda$.
In other words, this creates a feedback channel through which the covariate model can push the response model towards stronger smoothing than would be justified by $y$ alone.
If $f$ becomes too smooth, it can push spatial signal into $\beta$ and re-introduce bias.
This is an instance of Bayesian feedback \citep[e.g.,][]{mccandless2009bayesian, stephens2023causal}, where information flows between submodels through shared parameters.
Figure~\ref{fig:cut-dag2} visualizes these relationships in a graphical model.
\begin{figure}[t]
\centering
\begin{tikzpicture}[
		scale=0.8, transform shape,
		>=Latex,
		deterministic/.style={draw, fill=black!0, minimum width=10mm, minimum height=8mm},
		rv/.style={draw, circle, fill=black!10,minimum size=12mm},
		observed/.style={draw, circle, dashed, fill=blue!10, minimum size=12mm},
		strong/.style={->,line width=1pt,black!60},
		weak/.style={->,line width=1pt,orange!80},
		cutarrow/.style={->,dashed,red!70!black,very thick}
	]

	\node[rv] (sigma) at (1,3) {$\sigma$};
	\node[rv] (tau) at (4,2) {$\tau$};
	\node[rv] (xi_x) at (8,3) {$\xi_x$};
	\node[rv] (sigma_x) at (10,3) {$\sigma_x$};

	\node[deterministic] (tau_x) at (8,1) {$\tau_x$};
	
	\node[rv] (gamma_unpen) at (2.5, 0) {$\bm{\gamma}_{\mathrm{unpen}}$};
	\node[rv] (gamma_pen) at (4,0) {$\bm{\gamma}_{\mathrm{pen}}$};
	\node[rv] (beta) at (5.5, 0) {$\beta$};
	\node[rv] (gamma_x_pen) at (8,-0.5) {$\bm{\gamma}^x_{\mathrm{pen}}$};

	\node[deterministic] (loc_x) at (8,-2) {$\bm{f}^x$};
	\node[deterministic] (eps_x) at (6,-3) {$\bm{\varepsilon}_x$};
	\node[deterministic] (loc_y) at (4,-2) {$\bm{\mu}$};
	\node[rv] (gamma_x_unpen) at (9.5,-0.5) {\footnotesize $\bm{\gamma}^x_{\mathrm{unpen}}$};

	\node[observed] (y) at (2,-4) {$\bm{y}$};
	\node[observed] (x) at (8.7,-4) {$\bm{x}$};

	\draw[strong] (sigma) -- (y);
	\draw[strong] (tau) -- (gamma_pen);
	\draw[strong] (tau_x) -- (gamma_x_pen);
	\draw[strong] (x) -- (eps_x);
	\draw[strong] (loc_x) -- (eps_x);
	\draw[strong] (eps_x) -- (loc_y);
	\draw[strong] (gamma_pen) -- (loc_y);
	\draw[strong] (loc_y) -- (y);
	\draw[strong] (gamma_x_pen) -- (loc_x);
	\draw[strong] (gamma_unpen) -- (loc_y);
	\draw[strong] (gamma_x_unpen) -- (loc_x);
	\draw[strong] (beta) -- (loc_y);
	\draw[strong] (loc_x) -- (x);
	\draw[strong] (xi_x) -- (tau_x);
	\draw[strong] (sigma_x) -- (tau_x);
	\draw[strong] (sigma_x) to[bend left=23] (x);
	
	\draw[strong] (sigma) to[bend left=20] node[pos=0.75]{\red{\textbf{//}}} (tau_x) ; %
	\draw[strong] (tau) -- (tau_x) node[midway]{\red{\textbf{//}}};

\end{tikzpicture}

\caption{%
Graphical model representation of the Bayesian spatial+ model.
Random variables appear as gray circles with incoming arrows indicating conditional dependencies.
The two pale blue dashed circles at the bottom represent the observed data $y$ and $x$.
Boxes denote deterministic transformations, with incoming arrows indicating the variables they depend on.
For example, $\bm{\mu} = \bm{x}\beta + \bm{V}\bm{\gamma}_\text{pen} + \bm{U}\bm{\gamma}_\text{unpen}$.
The graphical structure reveals a problematic feedback channel from the covariate model to the response model as the prior for $\bm{\gamma}^x_{\text{pen}}$ depends on $\tau$ and $\sigma$ through $\tau_x$.
We eliminate this feedback (red bars indicate cut connections) by treating $\tau_x$ as fixed when updating $(\sigma, \tau)$, effectively removing the influence of $p(\bm{\gamma}^x_{\text{pen}} \mid \tau_x)$ on these parameters.
\label{fig:cut-dag2}}

\end{figure}

Our approach is to cut feedback \citep{mccandless2009bayesian} by preventing the covariate model from updating the response-side smoothness-related parameters $(\sigma, \tau)$. 
Conceptually, we retain the joint prior structure that ensures $\lambda_x < \lambda$, but we block the contribution of the covariate model to inference on $(\sigma, \tau)$.
Figure~\ref{fig:cut-dag2} makes this dependency entanglement explicit in terms of a graphical model representation.
Arrows indicate the dependence of $\bm{\gamma}^x_\text{pen}$ on $(\sigma, \tau)$ via the deterministic transformation $\tau_x$.
For inference on $(\sigma, \tau)$, we treat $\tau_x$ as fixed, which corresponds to removing the arrows from $\sigma$ and $\tau$ to $\tau_x$ and thus removing the feedback from $\bm{\gamma}^x_{\text{pen}}$ to those parameters, so that the posterior for $(\sigma,\tau)$ is driven by the response model alone. 
This preserves Bayesian coherence for all parameters except $(\sigma, \tau)$, but removes the problematic feedback channel and avoids oversmoothing driven by $x$.
We argue that this sacrifice of coherence is justified since our goal is not to find the optimal model for $x$ but rather ensure that the problematic, i.e. shared, frequencies are removed from it.

Our technique diverges from a traditional cut-feedback model -- typically motivated by causal inference. In our model, $(\tau, \sigma)$ are the only parameters in the prior hierarchy of $y$ that also appear in the hierarchy of $x$.
A traditional cut-feedback approach would discard all information from the response model when estimating $(\sigma, \tau)$, whereas we do the reverse by blocking information from the covariate model. Since $(\tau, \sigma)$ appear in the model for $x$ only through the derived parameter $\tau_x^2 = \tau^2 \frac{\sigma_x^2}{\sigma^2}\bigl(1+\xi_x^2\bigr)$, these hyperparameters become unidentifiable under the traditional approach.
In Supplement~\ref{app:trad-cut}, we confirm that the traditional cut-feedback approach does not work in our model: it produces extremely large standard errors (even on the log-scale) and systematically achieves coverage rates of 1, confirming the severe identifiability issues inherent in this approach.

\subsection{Estimation and Computational Details}
\label{sec:bsp-estimation2}
We perform Bayesian inference using Markov chain Monte Carlo (MCMC).
The unobserved model parameters comprise
\[
  \beta,
  \bm{\gamma}_{\mathrm{pen}}, \bm{\gamma}_{\mathrm{unpen}},
  \bm{\gamma}^x_{\mathrm{pen}}, \bm{\gamma}^x_{\mathrm{unpen}},
  \sigma, \sigma_x, \tau,  \xi_x
\]
and, in the prior-bounded model, we have $\lambda_{\mathrm{sp}}$ and $\xi_\lambda$ instead of $\tau$.
We employ a Metropolis-Hastings within Gibbs scheme.
The parameters blocks comprise of a block of spline and regression coefficients of the response model $(\beta, \bm{\gamma}_{\text{pen}}, \bm{\gamma}_{\text{unpen}})$, a block spline coefficients of the covariance model $(\bm{\gamma}^x_{\text{pen}}, \bm{\gamma}^x_{\text{unpen}})$ and a block of hyperparmeters $(\sigma, \sigma_x, \tau, \xi_x, \lambda_{\mathrm{sp}})$.
For the coefficient blocks, we use iteratively weighted least squares (IWLS) proposals \citep{gamerman1997sampling}.
Hyperparameters are updated on unrestricted domains (logarithmic transformation for positive parameters) using Hamiltonian Monte Carlo \citep[HMC, ][]{neal2011mcmc}.
This blocking strategy and use of IWLS proposals is following practices established in software for Bayesian regression such as BayesX \citep{brezger2005bayesx}.

All models are implemented in Python using \texttt{liesel} \citep{riebl2022liesel}, a probabilistic programming framework based on JAX and BlackJAX \citep{jax, cabezas2024blackjax}.
Besides full control over the model specification, the framework provides the flexibility needed for model specific MCMC algorithms such as the
cut-feedback mechanism.
Such customization would be challenging or impossible to implement in many standard frameworks such as STAN \citep{stan}.
We assess convergence using standard diagnostics such as the $\widehat{R}$ statistic and effective sample sizes \citep{gelman2006data, vehtari2021rank}.

We place half-normal priors on the hyperparameters. 
These are weakly informative shrinkage priors that keep hyperparameters close to zero when the data are weak, while avoiding the pathological posterior curvature that can arise from overly diffuse or heavy-tailed priors. T
his regularizes the posterior and leads to more stable Hamiltonian Monte Carlo geometry. 
In addition, the half-normal distribution has support on the positive reals, which is appropriate for these scale parameters.
Throughout the numerical experiments, we set all scale parameters in the priors on the hyperparameters to 10, except $\theta_{\xi_x} = 1$ yielding stronger regularization toward zero for $\xi_x$.
This tighter prior on $\xi_x$ ensures that $f^x$ does not become overly flexible and $\varepsilon^x$ retains essential high-frequency information, while the model setup always maintains the ordering constraint $\lambda_x < \lambda$. 
These hyperparameter values assume that parameters are approximately on unit scale after centering \citep[see][for a more detailed discussion of priors for variance parameters]{gelman2006prior, gelman2013bayesian}.

To model the spatial effects $f$ and $f_x$, we use thin plate regression splines, a low-rank approximation of thin plate splines \citep[][see Section~\ref{sec:ext-tprs}]{wood2003thin}. 
Additional details on MCMC diagnostics are provided alongside the simulations and applications.
The complete model hierarchy is shown in Supplement~\ref{app:hierarchy}.

\subsection{Variants and Extensions}
\label{sec:bsp-extensions}

This section presents extensions and variants of our methodology.
Initially, we discuss thin plate regression splines, which we use for computational efficiency in the simulations and applications.
Then, we introduce an alternative to cutting feedback when coherence is a primary concern and extend the model to multiple covariates.

\subsubsection{Thin plate regression splines}\label{sec:ext-tprs}

For computational efficiency, we use thin plate regression splines \citep[TPRS, ][]{wood2003thin}, which approximate the corresponding thin plate splines \citep{engle1986semiparametric, wahba1984partial} by the best rank-$k$ approximation in the spectral sense.
In a nutshell, this means that the regression spline basis is obtained from a spectral decomposition of the thin plate spline kernel at the data locations, retaining only $k$ eigencomponents corresponding to the smoothest (low-frequency) directions, together with low-order polynomial functions (e.g. linear functions of the input coordinates).
Additionally, the penalty is constructed from a roughness based penalty and includes in its null space the low-order polynomial component, which induces rank deficiency in the penalty matrix.

\subsubsection{Prior-bounded $\lambda$}\label{sec:ext-prior-bounded}

If coherence is of higher concern, we introduce an alternative to cutting feedback using Empirical Bayes.
We consider bounding the response-side smoothing from above by using an informative prior effectively limiting $\lambda$.
We introduce an upper bound $\lambda_{\mathrm{sp}} > 0$ and parameterize
$$
  \tau^2 = \sigma^2 \left(\frac{1}{\lambda_{\mathrm{sp}}} + \xi_\lambda^2\right),
  \qquad
  \xi_\lambda \sim \mathrm{HN}(0, \theta_\tau^2),
$$
which enforces $\lambda = \sigma^2/\tau^2 < \lambda_{\mathrm{sp}}$ while allowing less shrinkage through $\xi_\lambda^2$.
When sufficient prior knowledge is present, fixing the bound $\lambda_{\text{sp}}$ is an option, otherwise $\lambda_{\mathrm{sp}}$ is itself treated as a random quantity with an (weakly) informative prior.
The bound $\lambda_{\mathrm{sp}}$ is calibrated using posterior inference from a standard spatial model (see  Supplement~\ref{app:prior-bound-inference} for concrete details).
This prior-bounded version maintains a coherent joint posterior and limits oversmoothing by restricting how large $\lambda$ may become. The complete model hierarchy is shown in Supplement~\ref{app:hierarchy}.
In our empirical studies, both strategies reduce oversmoothing and bias, with the cut-feedback model typically providing the most robust performance.

\subsubsection{Multiple covariates}\label{sec:ext-multiple}
The methodology developed above naturally extends to multiple spatially-varying covariates.
Consider $p$ covariates $x_q(\bm{s})$, $q = 1, \dots, p$.
We naturally extend \eqref{eq:bayes_spatialplus_y} to
$$
y(\bm{s}_i) = \sum_{q=1}^p \beta_q \varepsilon^x_{q}(\bm{s}_i) + f(\bm{s}_i) + \varepsilon(\bm{s}_i)
$$
where the residual of each covariate model is created as in \eqref{eq:bayes_spatialplus_x}, i.e., $x_q(\bm{s}_i) =  f^{x}_{q}(\bm{s}_i) + \varepsilon^{x}_{q}(\bm{s}_i)$ with independent priors on the residual variance.

Following the preceding sections, we use common basis functions for all spatial effects and perform the reparameterization on each and place the priors as described on the penalized coefficient with variance parameters $\tau_{x, q}^2$.
Now, similarly to Equation~\eqref{eq:taux-param2}  for each $q$ we impose
\begin{align}
	\tau_{x,q}^2  &=  \tau^2 \frac{\sigma_{x,q}^2}{\sigma^2}(1  + \xi_{x,q}^2) \label{eq:tau_x_q},\ \mathrm{with}\ \xi_{x,q} \sim \mathrm{HN}(0,{\theta_x^q}^2), q = 1, \ldots, p.
\end{align}
The prior specification ensures that each covariate's spatial effect smoothness $\lambda_{x, q} = \sigma^2_{x, q}/\tau^2_{x, q}$ satisfies the smoothness restriction $\lambda_{x, q} < \lambda$, preventing that any covariate residual $\varepsilon^{x}_{q}$ contains frequency patterns shared with $f$.
The covariate-specific parameters driving $\tau_{x,q}^2$ are independent a-priori, allowing flexible smoothness control across covariates.
Control over Bayesian feedback is achieved by the natural extension of Section~\ref{sec:feedback2} to the new setting. Similarly, for the prior-bound variant.

When dealing with $p$ spatially aligned covariates, collinearity among covariates becomes a natural concern.
While not the primary focus of our here, we recommend preprocessing with standard techniques such as principal component analysis or combining highly collinear covariates to reduce collinearity before model fitting \citep[see Chapter~3 of][for details]{fahrmeir2001bayesian}.
Once these preprocessing steps are completed, the methods proposed in this section can be applied more effectively.
By reducing collinearity, the remaining covariates are expected to retain sufficient unique non-spatial or high-frequency spatial information, ensuring that each $\varepsilon^{x}_{q}$, $q=1, \dots, p$, retain enough information to accurately recover the associated coefficient estimate $\beta_q$.
An assessment of the effect of collinearity is provided in the simulation study in Section~\ref{sec:ss2}, with successful results for our proposed model.

\section{Simulation studies}\label{sec:ss}

In this section, we demonstrate the performance of Bayesian spatial+ over the frequentist counterpart using synthetic data, by evaluating performance in terms of bias reduction, coverage rates, and large-sample behavior. 
Section~\ref{sec:ss1} considers the case of one covariate and Section~\ref{sec:ss2} extends to two covariates.
Initially, we consider one covariate and then extend to two covariates.
The data setups are inspired by \citet{dupont2022spatial+}, \citet{dupont2023demystifying} and \citet{dupont2023spatial}.

We compare five models: a standard Bayesian spatial model (\texttt{spatial}), the frequentist spatial+ (\texttt{spatial+freq}), the Bayesian spatial+ without feedback control (\texttt{spatial+simple}), the prior-bounded Bayesian spatial+ variant (\texttt{spatial+hn}) from Section~\ref{sec:ext-prior-bounded}, and the Bayesian spatial+ with cut feedback (\texttt{spatial+cut}).  We do not fit any non-spatial models, as these are misspecified when there is unobserved spatial variability in the data, which is the case for both scenarios \citep{gilbert2025consistency}.

All models use TPRS with 500 basis functions. A pragmatic approach to choose $k$ is to increase $k$ if the estimated degrees of freedom for a TPRS exceeds some specified proportion of the basis dimension, commonly $80\%$ \citep[see][]{wood2003thin}. Frequentist models are fitted using \texttt{R-mgcv} \citep{wood2016just} and Bayesian models using \texttt{Liesel} \citep{riebl2022liesel}.
We use four MCMC chains, each with 2000 posterior samples and 3000 warm-up samples, and check the R-hat values based on MCMC chains for convergence \citep{gelman7dbrubin, vehtari2021rank}.

\subsection{One covariate}\label{sec:ss1}

We analyze spatial model performance across three scenarios using data at $n \in \{500, 1000\}$ locations sampled via Latin hypercube in $\mathcal{S} = [0, 1]^2$.
For 100 replicates, we generate
\begin{align*}
	y(\bm{s}_i) &= x(\bm{s}_i) + z_{high}(\bm{s}_i) + z_{low}(\bm{s}_i) + \varepsilon(\bm{s}_i), \\
	x(\bm{s}_i) &= z(\bm{s}_i) + \varepsilon^x(\bm{s}_i)
\end{align*}
where $\varepsilon(\bm{s}_i) \sim N(0, 0.1^2)$ and  $\varepsilon^x(\bm{s}_i) \sim N(0, \sigma_x^2)$.
Moreover, $z_{high}$ and $z_{low}$ are high and low frequency fields corresponding to the fitted values of a TPRS model with  $500$ and $20$ basis functions, respectively, to a realization of a Gaussian process with exponential covariance function $C_{\exp}(h) = \exp(-(h/R^{\exp}))$ with unit variance, where $h$ is the Euclidean distance between two observation locations, $R^{\exp}$ (related to spatial range) is set to $0.3$.
Consider also the high frequency field $z_{high2}$ generated similar to $z_{high}$ but based an independent realization of the underlaying process.
The scenarios differ in terms of $z(\bm{s}_i)$ and $\sigma_x$ as specified in Table~\ref{tab:sim1-scenarios}.

\begin{table}[b]
\centering
\begin{tabular}{rcc}
	\toprule
	 & $z(\bm{s}_i)$ & $\sigma_x$\\
	\midrule
	Scenario 1 & $z_{high}(\bm{s}_i)$ &  $0.1$\\
	Scenario 2& $z_{high}(\bm{s}_i)$& $0.2$\\
	Scenario 3 & $0.8 z_{low}(\bm{s}_i) + 0.2 z_{high2}(\bm{s}_i)$ &  $0.1$\\
	\bottomrule
\end{tabular}
\caption{Specifications of the three simulation scenarios. }\label{tab:sim1-scenarios}
\end{table}

We expect Scenario~1 to lead to biased estimates of $\beta$ (true value = 1) due to shared high frequencies between $x$ and the response's spatial effects ($z_{high}$).
In Scenario~2, we increase the amount of non-spatial information through larger $\sigma_x$, which helps identify the effects and should reduce bias compared to Scenario~1. %
In Scenario~3, we expect minimal bias since confounding occurs only at low frequencies, which experience less penalization and thus cause less distortion in coefficient estimation.

\paragraph{Results}

\begin{figure}[tb]
        \centering
        \includegraphics[width=\textwidth]{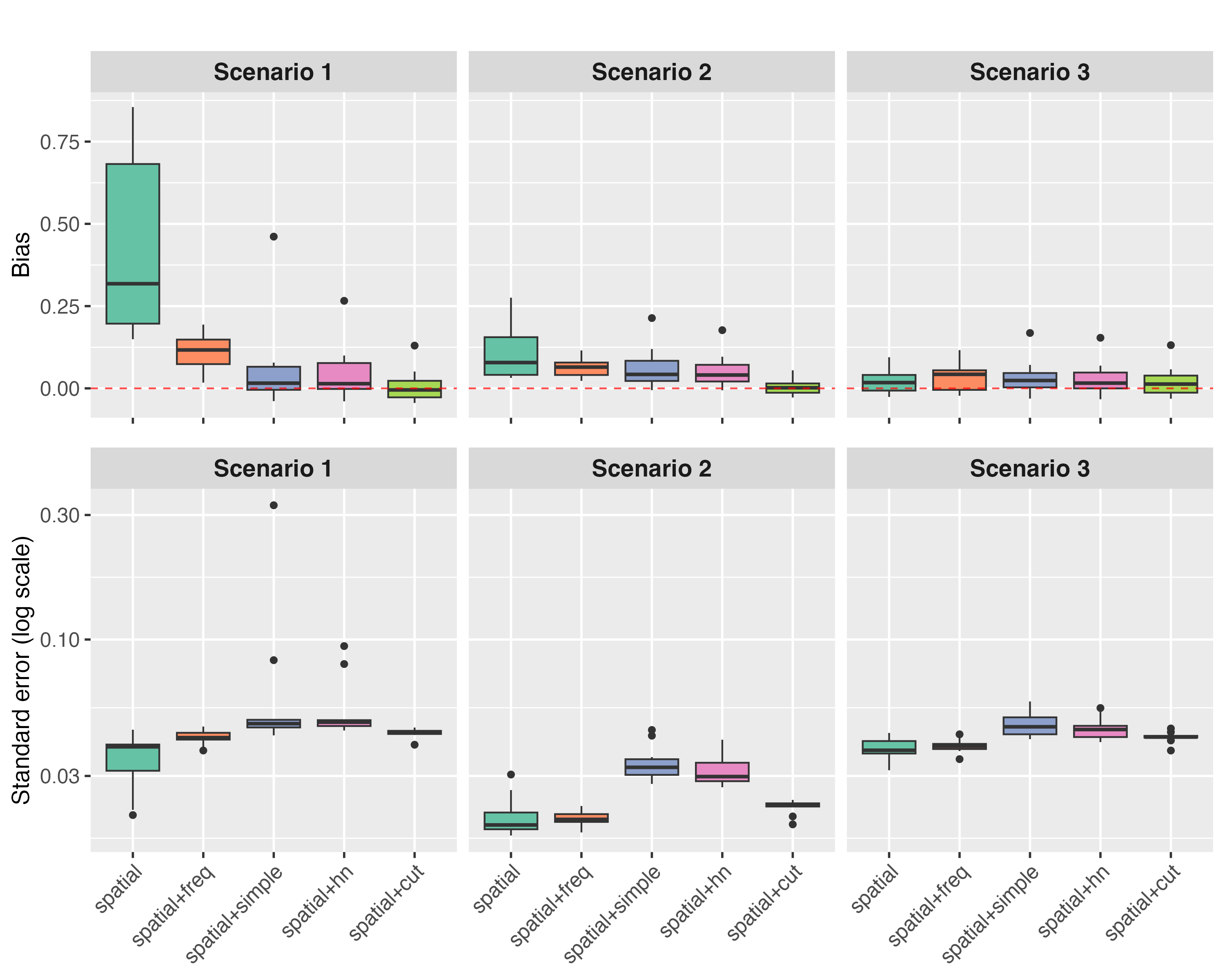}
    \caption{Bias estimate of $\beta$ (top) and standard error in log-scale (bottom) for Scenarios 1, 2 and 3 for $n = 1000$.}\label{fig:univariate_results}
\end{figure}

For $n=1000$, Figure~\ref{fig:univariate_results} shows the bias and standard deviation of the estimate of $\beta$ for the three scenarios. Table~\ref{tab:univariate} shows the coverage rates for $\beta$ based on 95\% confidence/credible intervals.
In Scenario~1, \texttt{spatial+cut} leads to the lowest bias and reaches the nominal coverage rate.
The next closest models are \texttt{spatial+simple} and \texttt{spatial+hn} with 75\% coverage, both outperforming the frequentist counterpart in terms of bias and coverage. The \texttt{spatial} model performs very poorly in both dimensions.
Regarding convergence diagnostics, \texttt{spatial+simple} struggles with reaching the threshold R-hat values compared to \texttt{spatial+hn} and \texttt{spatial+cut}. As expected, increasing $\sigma_x$ (Scenario~2) reduces bias in general.
This scenario also clarifies the need to cut feedback since $\lambda_x$ is pushed toward higher values as $\sigma_x$ increases, potentially affecting $\lambda$ via the constraint $\lambda_x <\lambda$.
Both \texttt{spatial+hn} and \texttt{spatial+cut} achieve better coverage rates than \texttt{spatial+simple}, with \texttt{spatial+cut} proving more effective than \texttt{spatial+hn} in handling feedback. 
In Scenario~3, all models reach bias close to zero and have high coverage rates. The frequentist spatial+ has a more noticeable positive bias, which could be explained by undersmoothing in the first stage that leads to residuals that are not very informative.

\begin{figure}[tb]
        \centering
        \includegraphics[width=\textwidth]{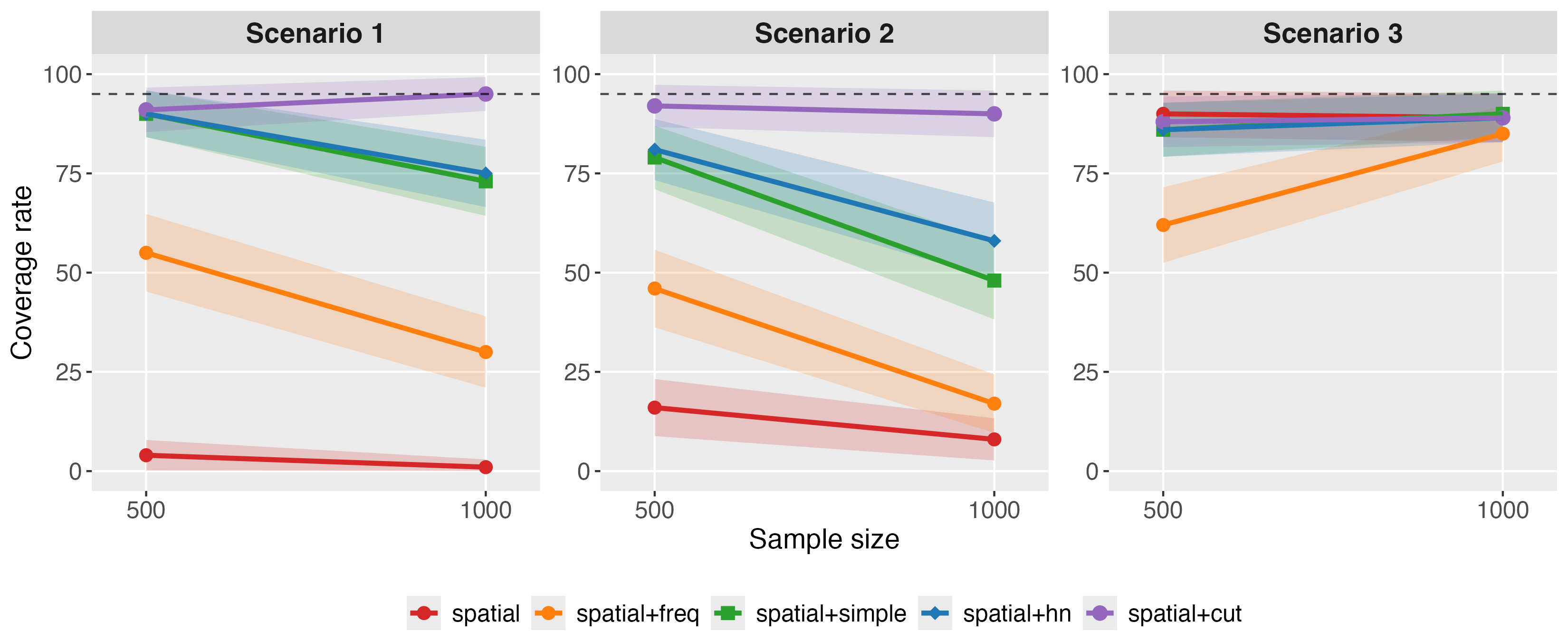}
    \caption{Coverage rates of intervals for one covariate as sample size increases from $n=500$ to $n=1000$. The shaded band represents the 95\% confidence interval for the empirical coverage estimate, computed using a normal approximation with variance derived from the binomial distribution.}\label{fig:large-sample}
\end{figure}

Results for $n=500$ show similar bias patterns (see Supplement~\ref{app:ss_500}), though coverage rates are generally closer to the nominal level in the confounded scenarios. 
Figure~\ref{fig:large-sample} shows coverage rates as sample size increases.
Notably, \texttt{spatial+cut} is the only method whose coverage is maintained with increasing $n$ when confounding is present and is overall close to the nominal level.
In contrast, other methods (except in the unconfounded scenario) exhibit deteriorating coverage as sample size grows.
A typical pattern of biased estimators, where larger samples yield tighter intervals centered on biased estimates (see Supplement~\ref{app:coverage}).
This finding further supports the advantage of \texttt{spatial+cut}.

\begin{table}[tb]
	\centering
	\medskip
	\begin{tabular}{lccccc}
	\toprule
	 & \texttt{spatial} & \texttt{spatial+freq} & \texttt{spatial+simple} & \texttt{spatial+hn} & \texttt{spatial+cut} \\
	\midrule
	Scenario 1 & 0.01 & 0.30 & 0.73 & 0.75 & 0.95 \\
	Scenario 2 & 0.08 & 0.17 & 0.48 & 0.58 & 0.90 \\
	Scenario 3 & 0.89 & 0.85 & 0.90 & 0.89 & 0.89 \\
	\bottomrule
	\end{tabular}
	\caption{Coverage rates for $\beta$ based on 95\% intervals ($n=1000$).
	} \label{tab:univariate}
\end{table}

\subsection{Two covariates}\label{sec:ss2}

We assess model performance with two spatially aligned covariates using 100 replicates at $n=1000$ locations (Latin hypercube sampling as above).
Inspired by \citet{dupont2023spatial}, data is generated according to
\begin{align*}
	y(\bm{s}_i) &= 0.5 x_1(\bm{s}_i) + 0.5 x_2(\bm{s}_i) + z_{high}(\bm{s}_i) + z_{low}(\bm{s}_i) + \varepsilon(\bm{s}_i), \\
	x_1(\bm{s}_i) &= 0.5 z_{high}(\bm{s}_i) + 0.5 z_{high2}(\bm{s}_i) + \varepsilon^{x}_1(\bm{s}_i), \\
	x_2(\bm{s}_i) &= z(\bm{s}_i) + \varepsilon^{x}_2(\bm{s}_i)
\end{align*}
with $\varepsilon(\bm{s}_i) \sim N(0, 0.1^2)$ and  $\varepsilon^{x}_1(\bm{s}_i), \varepsilon^{x}_2(\bm{s}_i)  \sim N(0, 0.1^2)$.
The values of $z_{high}$, $z_{high2}$, and $z_{low}$ are generated analogously to the one covariate case.
The scenarios differ in terms how $z(\bm{s}_i)$ is defined. 
In Scenario~1, define $z(\bm{s}_i) = 0.8z_{high}(\bm{s}_i) +  0.2z_{high2}(\bm{s}_i)$; in Scenario~2, define $z(\bm{s}_i) = 0.2 z_{high}(\bm{s}_i) +  0.8z_{high2}(\bm{s}_i)$; in Scenario~3, define $z(\bm{s}_i) = 0.7 z_{low}(\bm{s}_i) + 0.1 z_{high}(\bm{s}_i)  + 0.2z_{high2}(\bm{s}_i)$.
The first scenario represents a setting where $x_2$ shares high frequencies with $f$ and $x_1$ with a relatively large weight.
In the second scenario, this weight drops to 0.2 for $x_2$, such that we would expect generally smaller bias in $\beta_2$.
In Scenario~3, $x_1$ and $x_2$ are again not strongly correlated; $f$ and $x_2$ share low frequencies through $z_{low}$ with a weight of 0.7, but shared low frequencies lead to little to no spatial confounding.

\begin{figure}[tb]
        \centering
        \includegraphics[width=\textwidth]{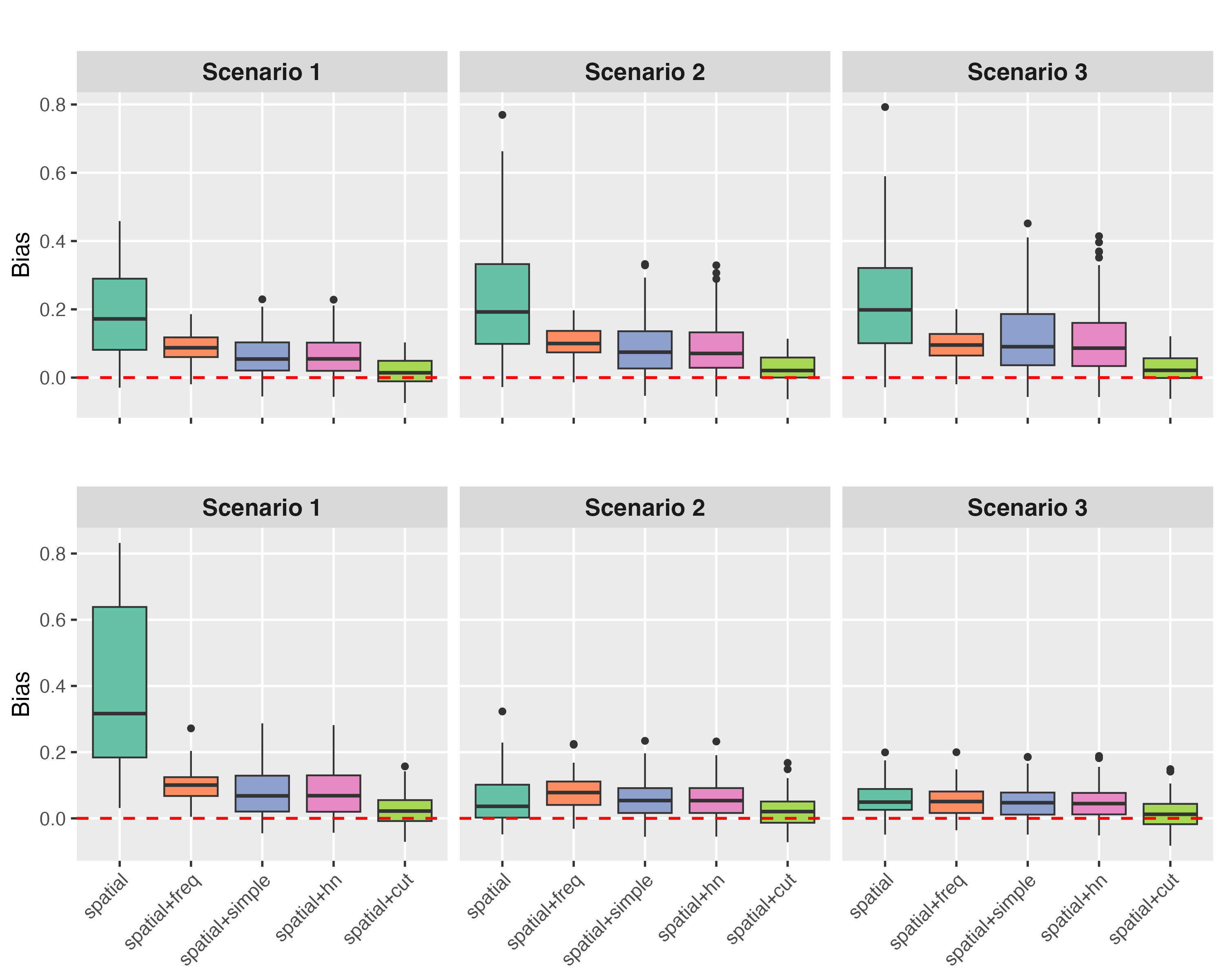}
    \caption{Bias estimate of $\beta_1$ (top) and $\beta_2$ (bottom) for Scenarios 1, 2 and 3.}\label{fig:multiple_results}
\end{figure}

\paragraph{Results}
Figure~\ref{fig:multiple_results} and Table~\ref{tab:multiple2} present the bias and coverage results for both coefficients $\beta_1$ and $\beta_2$ (associated with $x_1$ and $x_2$, respectively) across all three scenarios.
\texttt{spatial+cut} consistently achieves the lowest median bias and highest coverage rates across all scenarios and coefficients, substantially outperforming the frequentist approach and standard spatial model. The estimates for $\beta_1$ remain stable across scenarios since $x_1$ maintains the same structure throughout, with optimal performance in Scenario~1 where both covariates are more noticeably confounded amongst themselves.
In Scenario~2, $x_2$ suffers less from spatial confounding and most models improve, with the notable exception of \texttt{spatial+freq}, which exhibits similar bias to Scenario~1. This may be because the frequentist approach removes too much information from $\varepsilon^x$, making it difficult to accurately identify $\beta_2$, or is not able to avoid frequency leakage from the first-stage.
In Scenario~3, $\beta_2$ exhibits low bias across all models, as expected given the minimal spatial confounding in this setting. Still, the Bayesian spatial+ models do significantly better in terms of coverage of $\beta_2$ than the remaining models.
Supplement~\ref{app:ss_multiple} presents the standard errors associated with the estimates, showing that \texttt{spatial+freq} and \texttt{spatial+cut} achieve similar precision for the coefficient estimates, while \texttt{spatial+simple} and \texttt{spatial+hn} exhibit the largest standard errors.

\begin{table}[tb]
 \begin{minipage}{.5\linewidth}
	\centering
	\begin{tabular}{lccc}
	\multicolumn{4}{c}{$\beta_1$}\\
	\toprule
	& S1 & S2 & S3\\
	\midrule
	\texttt{spatial} & 0.15 & 0.12 & 0.10\\
	\texttt{spatial+freq} & 0.44 & 0.32 & 0.38\\
	\texttt{spatial+simple} & 0.79 & 0.61 & 0.60\\
	\texttt{spatial+hn} & 0.78 & 0.62 & 0.61\\
	\texttt{spatial+cut} & 0.94 & 0.90 & 0.90\\
	\bottomrule
	\end{tabular}
    \end{minipage}%
    \begin{minipage}{.5\linewidth}
	\centering
	\begin{tabular}{l c c c}
	\multicolumn{4}{c}{$\beta_2$}\\
	\toprule
	& S1 & S2 & S3\\
	\midrule
	\texttt{spatial} & 0.03 & 0.64 & 0.68\\
	\texttt{spatial+freq} & 0.42 & 0.55 & 0.73\\
	\texttt{spatial+simple} & 0.65 & 0.80 & 0.91\\
	\texttt{spatial+hn} & 0.66 & 0.80 & 0.91\\
	\texttt{spatial+cut} & 0.93 & 0.90 & 0.95\\
	\bottomrule
	\end{tabular}
	\end{minipage}
	\caption{Coverage rates for $\beta_1$ (left table) and  $\beta_2$ (right table) based on 95\% confidence/credible intervals ($n=1000$). S1, S2 and S3 refer to Scenarios 1, 2 and 3.} \label{tab:multiple2}
\end{table}

\section{Applications to real data}\label{sec:application}

\subsection{Average daily precipitation in Germany}\label{sec:application1}
\begin{figure}[tb]
        \centering
        \includegraphics[width=\textwidth]{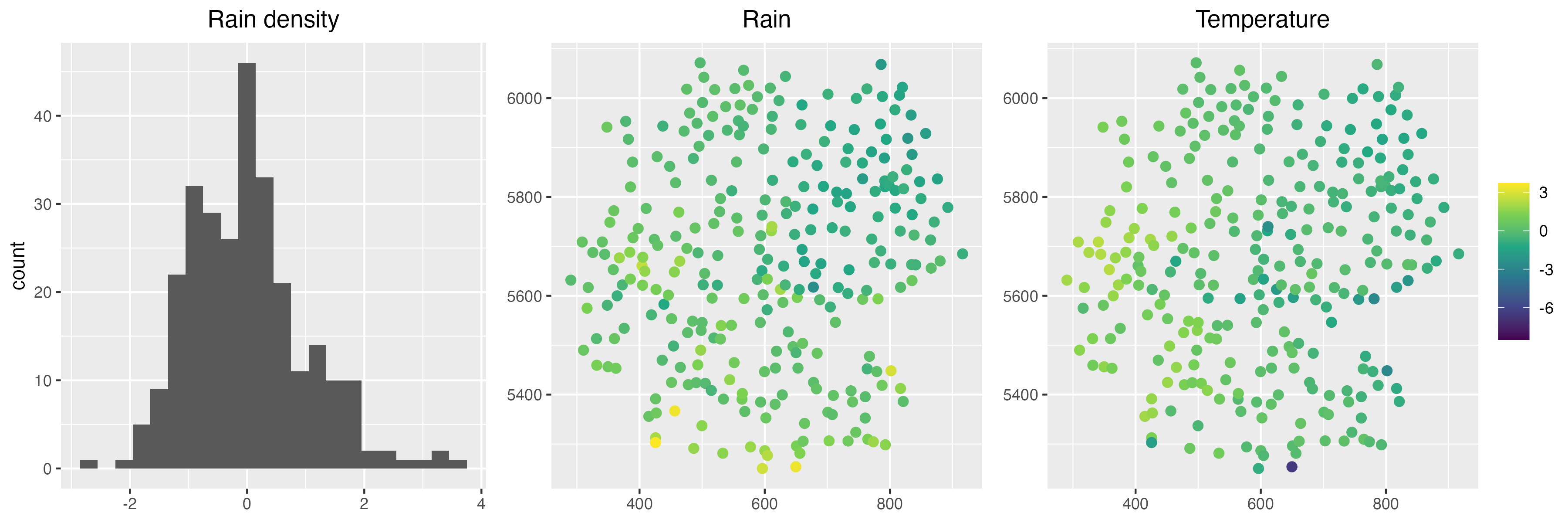}
    \caption{Histogram for the response rain (left); scatterplots for the transformed response, rainfall (center), and the covariate, air temperature (right).}\label{fig:rain}
\end{figure}

We analyze the average amount of daily precipitation in February 2016 across Germany, using air temperature as a covariate.
The dataset includes observations from 279 weather stations.
We apply a square-root transformation to the rainfall variable to stabilize variance and then standardize all variables.
Figure~\ref{fig:rain} illustrates the transformed and standardized variables.
The scatterplots reveal a common smaller-scale spatial pattern between rainfall and air temperature.
Specifically, lower temperatures are observed from the east to the central and southern regions of the country, coinciding with locations with larger amounts of precipitation.
This pattern aligns with areas of higher altitude.
Given the strong correlation between temperature and altitude, we chose not to include both variables simultaneously in the model.

\paragraph{Results}
We fit all models to the data using TPRS with 150 basis functions and use four MCMC chains, each with 2000~posterior samples and a warm-up comprised of 3000~iterations.
Table~\ref{tab:application} presents the confidence and credible intervals for the estimate of $\beta_{temp}$.
The results indicate evidence of spatial confounding since the estimates of the models corrected for spatial confounding change compared to the spatial model.
Relative to the standard spatial model ($\hat{\beta}_{temp}=-0.82$), \texttt{spatial+freq} shows increased magnitude ($\hat{\beta}_{temp}=-0.89$), while the Bayesian spatial+ methods achieve larger corrections ($\hat{\beta}_{temp} = -0.97$).
The consistency across Bayesian spatial+ methods suggests robust bias correction, with \texttt{spatial+cut} performing similarly to \texttt{spatial+hn} and \texttt{spatial+simple}.
These findings align with our simulation results, confirming that Bayesian spatial+ methods provide more effective spatial confounding correction than the frequentist approach.
An $\hat{R} < 1.1$ indicates convergence.

\begin{table}[b]
	\centering
	\begin{tabular}{lccc}
	\toprule
	 & $\hat{\beta}_{temp}$ & $\operatorname{se}(\hat{\beta}_{temp})$  & $\operatorname{CI}(\beta_{temp})$  \\
	\midrule
    \texttt{spatial}  &  -0.82 & 0.0526 & [-0.93, -0.71] \\
    \texttt{spatial+freq} &   -0.89 &  0.0663 & [-1.02, -0.76] \\
   	\texttt{spatial+simple} & -0.97 &  0.0625  & [-1.09, -0.85]   \\
    \texttt{spatial+hn} & -0.97 &  0.0611 & [-1.09, -0.86]  \\
    \texttt{spatial+cut} & -0.97 &  0.0613 & [-1.08, -0.85]  \\
   \bottomrule
	\end{tabular}
	\caption{Point estimate $\hat{\beta}_{temp}$ and corresponding standard error (se) and confidence/credibility intervals (CI) for all models.}\label{tab:application}
\end{table}

\subsection{Tree coverage in Germany}\label{sec:application2}
\begin{figure}[tb]
        \centering
        \includegraphics[width=\textwidth]{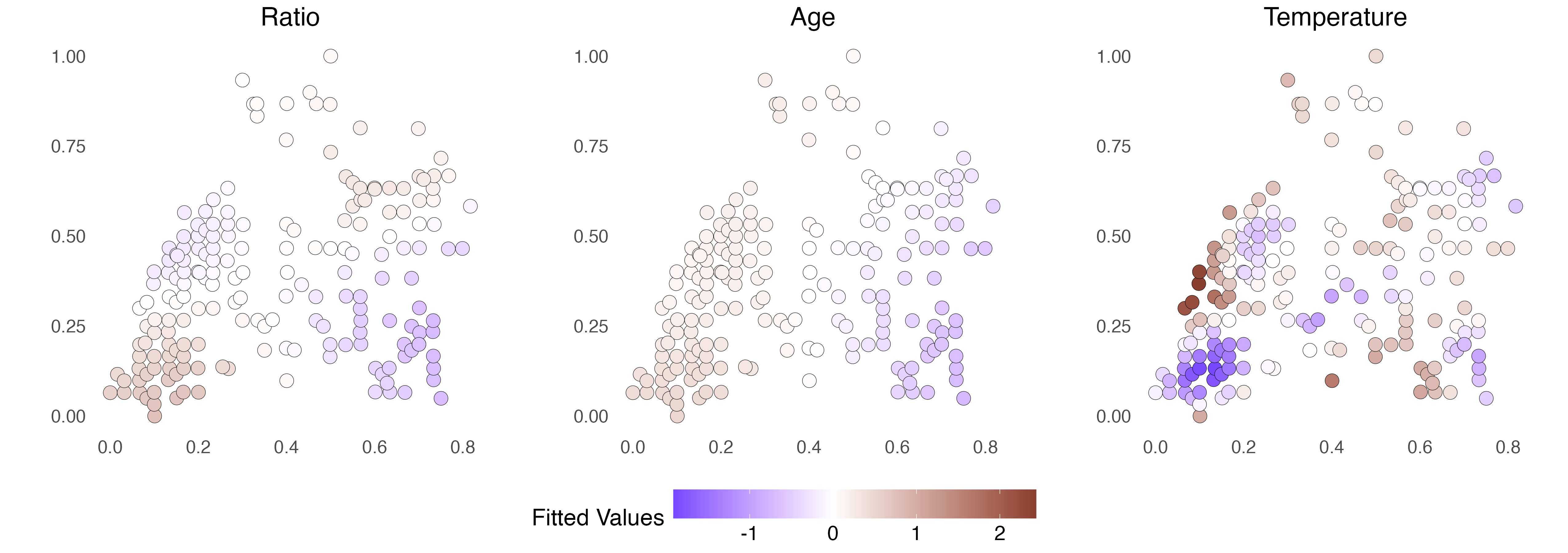}  
    \caption{We fit thin plate regression splines applied to defoliation (left), age (center), temperature (right) using \texttt{mgcv} and show the fitted values of the spatial smooth; $x$ and $y$ the spatial coordinates.}\label{fig:spatial_structure}
\end{figure}

We revisit an application in \citet[][data provided in the supplementary material]{dupont2022spatial+} and details on the original data are available at \citet{augustin2009modeling}.
The dataset considers data on spruce trees collected in 2013, corresponding to $n = 186$ observation locations in Germany.
We study the effect of minimum temperature in May (temp) and tree age (age) on crown defoliation (defol) expressed as a ratio. 
High minimum temperature in May is indicative of a warmer year which, in turn, is likely to lead to higher levels of tree defoliation (measured later in the summer). Moreover, it is expected that older trees have more defoliation.
All variables are standardized.
Figure~\ref{fig:spatial_structure} displays the fitted values of fitting a smooth TPRS to defoliation (left), age (center), and temperature (right), independently.
The spatial components of defoliation and age are quite smooth and spatial TPRSs can only explain a small percentage of the deviances in the data (less than 13\%).
In contrast, the spatial structure of temperature appears considerably more complex and the TPRS explains about 76\% of the deviance in the data.

\paragraph{Results}

We fit the models using $100$ basis functions and for the Bayesian models run four MCMC chains with 2000 posterior samples and a warm-up of 3000 iterations.
Convergence is verified using $\hat{R}$.
Table~\ref{tab:application2} presents the estimated coefficient for the different models.
The coefficients for temperature approximately double in magnitude for the spatial+ models compared to the standard spatial model, while the coefficients for age are estimated consistently across all models.
The results, consistent with \citet{dupont2022spatial+}, indicate that age is not affected by spatial confounding, but temperature is.
The Bayesian and frequentist versions of the spatial+ models give similar results, with \texttt{spatial+cut} having the smallest coefficient estimate and \texttt{spatial+freq} having the largest.
The cut feedback method has a 8\% smaller standard error compared with the frequentist counter part.

\begin{table}[tb]
	\centering
	\begin{tabular}{lcccccc}
	\toprule
	& $\hat{\beta}_{age}$ & $\operatorname{se}_{age}$  & $\operatorname{CI}(\hat{\beta}_{age})$ & $\hat{\beta}_{temp}$ & $\operatorname{se}(\hat{\beta}_{temp})$ & $\operatorname{CI}(\beta_{temp})$ \\
	\midrule
	\texttt{spatial}  &  0.6787 & 0.0537 & [0.5741, 0.7836] & 0.1199 & 0.0602 & [0.0022, 0.2381]\\
   	\texttt{spatial+freq} & 0.6749 & 0.0529 & [0.5704, 0.7793]  & 0.2694 & 0.0990 & [0.0740, 0.4648] \\
   	\texttt{spatial+simple} &  0.6854  & 0.0537 & [0.5810, 0.7915] & 0.2565 & 0.0981 & [0.0636, 0.4519]\\
    \texttt{spatial+hn} &  0.6855 & 0.0535 & [0.5801, 0.7909] & 0.2560 & 0.0957 & [0.0719, 0.4447]  \\
    \texttt{spatial+cut} &  0.6848 & 0.0563 & [0.5725, 0.7948] & 0.2487 & 0.0929 & [0.0655, 0.4301]  \\
    \bottomrule  
	\end{tabular}
	\caption{Point estimates $\hat{\beta}_{temp}$ and  $\hat{\beta}_{age}$ and the corresponding standard errors (se) and confidence/credibility intervals (CI) for all estimated models.}\label{tab:application2}
\end{table}

\section{Discussion}\label{sec:conclusion}
This work brings the spatial+ methodology to a Bayesian framework, offering three key advantages.
First, we provide enhanced uncertainty quantification  with seamless uncertainty propagation in the joint model.
Second, this enables robust posterior inference beyond point estimates, addressing a gap in the original spatial+ framework where inference beyond the mean was not explicitly detailed.
Third, the Bayesian approach allows incorporating prior knowledge about spatial confounding, enabling restrictions that explicitly avoid confounded high frequencies. %
Additionally, we identify a feedback problem that emerges specifically in the joint Bayesian formulation and propose a cutting feedback approach that differs from traditional methods by selectively removing problematic dependencies from parameters in the response side.
We extended the presented methodology to multiple covariates.

We evaluated our approach through simulation studies and applications to two real-world datasets with both single and multiple covariates. 
Results demonstrate that the Bayesian spatial+ models, particularly the cutting feedback model, represent a substantial improvement over the frequentist counterpart in both bias and coverage, and the spatial model performed consistently worse than all spatial+ models, except when there was no relevant confounding.
Importantly, our approach achieves these improvements without requiring user-specified tuning parameters such as frequency caps, making it both more effective and more practical than existing alternatives.
Nevertheless, the current requirement of separability between residuals and spatial effects presents challenges when extending spatial+ to non-Gaussian covariates.
Future research could focus on developing more flexible methodologies to address these limitations and further expand the applicability of Bayesian spatial+.

\section*{Acknowledgements}

We extend our gratitude to Thomas Kneib for helpful comments and discussions.
We also acknowledge the valuable feedback received from editors and anonymous reviewers.
IM acknowledges support by the Deutsche Forschungsgemeinschaft (DFG, German Research Foundation) through CRC~990 ``Ecological and Socioeconomic Functions of Tropical Lowland Rainforest Transformation Systems''. 

	\bibliography{mybib.bib}

\begin{thebibliography}{}

\bibitem[Augustin et~al., 2009]{augustin2009modeling}
Augustin, N.~H., Musio, M., von Wilpert, K., Kublin, E., Wood, S.~N., and
  Schumacher, M. (2009).
\newblock Modeling spatiotemporal forest health monitoring data.
\newblock {\em Journal of the American Statistical Association},
  104(487):899--911.

\bibitem[Bradbury et~al., 2024]{jax}
Bradbury, J., Frostig, R., Hawkins, P., Johnson, M.~J., and Leary, C. (2024).
\newblock {JAX: Autograd and XLA}.

\bibitem[Brezger et~al., 2005]{brezger2005bayesx}
Brezger, A., Kneib, T., and Lang, S. (2005).
\newblock Bayesx: analyzing bayesian structural additive regression models.
\newblock {\em Journal of Statistical Software}, 14:1--22.

\bibitem[Cabezas et~al., 2024]{cabezas2024blackjax}
Cabezas, A., Corenflos, A., Lao, J., Louf, R., Carnec, A., Chaudhari, K.,
  Cohn-Gordon, R., Coullon, J., Deng, W., Duffield, S., Durán-Martín, G.,
  Elantkowski, M., Foreman-Mackey, D., Gregori, M., Iguaran, C., Kumar, R.,
  Lysy, M., Murphy, K., Orduz, J.~C., Patel, K., Wang, X., and Zinkov, R.
  (2024).
\newblock {BlackJAX: Composable Bayesian inference in JAX}.

\bibitem[Carroll et~al., 2006]{carroll2006measurement}
Carroll, R.~J., Ruppert, D., Stefanski, L.~A., and Crainiceanu, C.~M. (2006).
\newblock {\em Measurement error in nonlinear models: a modern perspective}.
\newblock Chapman and Hall/CRC.

\bibitem[Clayton et~al., 1993]{clayton1993spatial}
Clayton, D.~G., Bernardinelli, L., and Montomoli, C. (1993).
\newblock Spatial correlation in ecological analysis.
\newblock {\em International Journal of Epidemiology}, 22(6):1193--1202.

\bibitem[Dupont and Augustin, 2023]{dupont2023spatial}
Dupont, E. and Augustin, N.~H. (2023).
\newblock Spatial confounding and spatial+ for nonlinear covariate effects.
\newblock {\em Journal of Agricultural, Biological and Environmental
  Statistics}, pages 1--16.

\bibitem[Dupont et~al., 2023]{dupont2023demystifying}
Dupont, E., Marques, I., and Kneib, T. (2023).
\newblock Demystifying spatial confounding.
\newblock {\em arXiv preprint arXiv:2309.16861}.

\bibitem[Dupont et~al., 2022]{dupont2022spatial+}
Dupont, E., Wood, S.~N., and Augustin, N.~H. (2022).
\newblock Spatial+: a novel approach to spatial confounding.
\newblock {\em Biometrics}, 78(4):1279--1290.

\bibitem[Engle et~al., 1986]{engle1986semiparametric}
Engle, R.~F., Granger, C.~W., Rice, J., and Weiss, A. (1986).
\newblock Semiparametric estimates of the relation between weather and
  electricity sales.
\newblock {\em Journal of the American Statistical Association},
  81(394):310--320.

\bibitem[Fahrmeir and Kneib, 2009]{fahrmeir2009propriety}
Fahrmeir, L. and Kneib, T. (2009).
\newblock Propriety of posteriors in structured additive regression models:
  Theory and empirical evidence.
\newblock {\em Journal of Statistical Planning and Inference}, 139(3):843--859.

\bibitem[Fahrmeir et~al., 2004]{fahrmeir2004penalized}
Fahrmeir, L., Kneib, T., and Lang, S. (2004).
\newblock Penalized structured additive regression for space-time data: a
  {Bayesian} perspective.
\newblock {\em Statistica Sinica}, 14(3):731--761.

\bibitem[Fahrmeir and Lang, 2001]{fahrmeir2001bayesian}
Fahrmeir, L. and Lang, S. (2001).
\newblock Bayesian semiparametric regression analysis of multicategorical
  time-space data.
\newblock {\em Annals of the Institute of Statistical Mathematics}, 53:11--30.

\bibitem[Gamerman, 1997]{gamerman1997sampling}
Gamerman, D. (1997).
\newblock Sampling from the posterior distribution in generalized linear mixed
  models.
\newblock {\em Statistics and Computing}, 7(1):57--68.

\bibitem[Gelman, 2006]{gelman2006prior}
Gelman, A. (2006).
\newblock Prior distributions for variance parameters in hierarchical models.
\newblock {\em Bayesian Analysis}, 1(3):515--534.

\bibitem[Gelman et~al., 1992]{gelman7dbrubin}
Gelman, A., Carlin, J., Stern, H., and Rubin, D. (1992).
\newblock Inference from iterative simulation using multiple sequences.
\newblock {\em Statistical Science}, 7(4):457--511.

\bibitem[Gelman and Hill, 2006]{gelman2006data}
Gelman, A. and Hill, J. (2006).
\newblock {\em Data analysis using regression and multilevel/hierarchical
  models}.
\newblock Cambridge University Press.

\bibitem[Gelman et~al., 2013]{gelman2013bayesian}
Gelman, A., Stern, H.~S., Carlin, J.~B., Dunson, D.~B., Vehtari, A., and Rubin,
  D.~B. (2013).
\newblock {\em Bayesian data analysis}.
\newblock Chapman and Hall/CRC.

\bibitem[Gelman and Tuerlinckx, 2000]{gelman2000type}
Gelman, A. and Tuerlinckx, F. (2000).
\newblock Type s error rates for classical and bayesian single and multiple
  comparison procedures.
\newblock {\em Computational Statistics}, 15(3):373--390.

\bibitem[Gilbert et~al., 2025]{gilbert2025consistency}
Gilbert, B., Ogburn, E.~L., and Datta, A. (2025).
\newblock Consistency of common spatial estimators under spatial confounding.
\newblock {\em Biometrika}, 112(2):asae070.

\bibitem[Guan et~al., 2022]{guan2022spectral}
Guan, Y., Page, G.~L., Reich, B.~J., Ventrucci, M., and Yang, S. (2022).
\newblock A spectral adjustment for spatial confounding.
\newblock {\em Biometrika}, 110(3):699--719.

\bibitem[Hanks et~al., 2015]{hanks2015restricted}
Hanks, E.~M., Schliep, E.~M., Hooten, M.~B., and Hoeting, J.~A. (2015).
\newblock Restricted spatial regression in practice: geostatistical models,
  confounding, and robustness under model misspecification.
\newblock {\em Environmetrics}, 26(4):243--254.

\bibitem[Khan and Calder, 2020]{khan2020restricted}
Khan, K. and Calder, C.~A. (2020).
\newblock Restricted spatial regression methods: implications for inference.
\newblock {\em Journal of the American Statistical Association},
  117(537):482--494.

\bibitem[Klein et~al., 2016]{klein2016scale}
Klein, N., Kneib, T., et~al. (2016).
\newblock Scale-dependent priors for variance parameters in structured additive
  distributional regression.
\newblock {\em Bayesian Analysis}, 11(4):1071--1106.

\bibitem[Klein et~al., 2015]{klein2015bayesian}
Klein, N., Kneib, T., Lang, S., Sohn, A., et~al. (2015).
\newblock Bayesian structured additive distributional regression with an
  application to regional income inequality in germany.
\newblock {\em The Annals of Applied Statistics}, 9(2):1024--1052.

\bibitem[M{\"a}kinen et~al., 2022]{makinen2022spatial}
M{\"a}kinen, J., Numminen, E., Niittynen, P., Luoto, M., and Vanhatalo, J.
  (2022).
\newblock Spatial confounding in {Bayesian} species distribution modeling.
\newblock {\em Ecography}, 2022(11):e06183.

\bibitem[Marques and Kneib, 2022]{marques2022spatial+}
Marques, I. and Kneib, T. (2022).
\newblock Discussion on 'spatial+: A novel approach to spatial confounding` by
  {Dupont, E., Wood, S. and Augustin, N.}
\newblock {\em Biometrics}.

\bibitem[Marques et~al., 2022]{marques2022mitigating}
Marques, I., Kneib, T., and Klein, N. (2022).
\newblock Mitigating spatial confounding by explicitly correlating {Gaussian}
  random fields.
\newblock {\em Environmetrics}, 33(5):e2727.

\bibitem[McCandless et~al., 2009]{mccandless2009bayesian}
McCandless, L.~C., Gustafson, P., and Austin, P.~C. (2009).
\newblock Bayesian propensity score analysis for observational data.
\newblock {\em Statistics in Medicine}, 28(1):94--112.

\bibitem[Neal, 2011]{neal2011mcmc}
Neal, R.~M. (2011).
\newblock Mcmc using hamiltonian dynamics.
\newblock In {\em Handbook of Markov Chain Monte Carlo}, volume~2, page~2. CRC
  press.

\bibitem[Nychka, 2000]{nychka2000spatial}
Nychka, D.~W. (2000).
\newblock Spatial-process estimates as smoothers.
\newblock In Schimek, M.~G., editor, {\em Smoothing and Regression: Approaches,
  Computation, and Application}. Wiley.

\bibitem[Paciorek, 2010]{paciorek2010importance}
Paciorek, C.~J. (2010).
\newblock The importance of scale for spatial-confounding bias and precision of
  spatial regression estimators.
\newblock {\em Statistical Science}, 25(1):107--125.

\bibitem[Ramsay, 2002]{ramsay2002spline}
Ramsay, T. (2002).
\newblock Spline smoothing over difficult regions.
\newblock {\em Journal of the Royal Statistical Society Series B: Statistical
  Methodology}, 64(2):307--319.

\bibitem[Reich et~al., 2006]{reich2006effects}
Reich, B.~J., Hodges, J.~S., and Zadnik, V. (2006).
\newblock Effects of residual smoothing on the posterior of the fixed effects
  in disease-mapping models.
\newblock {\em Biometrics}, 62(4):1197--1206.

\bibitem[Reich et~al., 2022]{reich2022spatial+}
Reich, B.~J., Yang, S., and Guan, Y. (2022).
\newblock Discussion on `spatial+: A novel approach to spatial confounding' by
  dupont, wood, and augustin.
\newblock {\em Biometrics}, 78(4):1291--1294.

\bibitem[Riebl et~al., 2022]{riebl2022liesel}
Riebl, H., Wiemann, P.~F., and Kneib, T. (2022).
\newblock Liesel: a probabilistic programming framework for developing
  semi-parametric regression models and custom {Bayesian} inference algorithms.
\newblock {\em arXiv preprint arXiv:2209.10975}.

\bibitem[Sangalli et~al., 2013]{sangalli2013spatial}
Sangalli, L.~M., Ramsay, J.~O., and Ramsay, T.~O. (2013).
\newblock Spatial spline regression models.
\newblock {\em Journal of the Royal Statistical Society Series B: Statistical
  Methodology}, 75(4):681--703.

\bibitem[Schmidt, 2022]{schmidt2022discussion}
Schmidt, A.~M. (2022).
\newblock Discussion on ``spatial+: A novel approach to spatial confounding''
  by emiko dupont, simon n. wood, and nicole h. augustin.
\newblock {\em Biometrics}, 78(4):1300--1304.

\bibitem[{Stan Development Team}, 2023]{stan}
{Stan Development Team} (2023).
\newblock Stan modeling language users guide and reference manual.
\newblock {\em https://mc-stan.org}.

\bibitem[Stephens et~al., 2023]{stephens2023causal}
Stephens, D.~A., Nobre, W.~S., Moodie, E.~E., and Schmidt, A.~M. (2023).
\newblock Causal inference under mis-specification: adjustment based on the
  propensity score (with discussion).
\newblock {\em Bayesian Analysis}, 18(2):639--694.

\bibitem[Thaden and Kneib, 2018]{thaden2018structural}
Thaden, H. and Kneib, T. (2018).
\newblock Structural equation models for dealing with spatial confounding.
\newblock {\em The American Statistician}, 72(3):239--252.

\bibitem[Ugarte et~al., 2017]{ugarte2017one}
Ugarte, M., Adin, A., and Goicoa, T. (2017).
\newblock One-dimensional, two-dimensional, and three dimensional {B}-splines
  to specify space--time interactions in {Bayesian} disease mapping: model
  fitting and model identifiability.
\newblock {\em Spatial Statistics}, 22:451--468.

\bibitem[Urdangarin et~al., 2024]{urdangarin2023one}
Urdangarin, A., Goicoa, T., Kneib, T., and Ugarte, M.~D. (2024).
\newblock A simplified spatial+ approach to mitigate spatial confounding in
  multivariate spatial areal models.
\newblock {\em Spatial Statistics}, 59:100804.

\bibitem[Urdangarin et~al., 2022]{urdangarin2022evaluating}
Urdangarin, A., Goicoa, T., and Ugarte, M.~D. (2022).
\newblock Evaluating recent methods to overcome spatial confounding.
\newblock {\em Revista Matem{\'a}tica Complutense}, 36:333--360.

\bibitem[Vehtari et~al., 2021]{vehtari2021rank}
Vehtari, A., Gelman, A., Simpson, D., Carpenter, B., and B{\"u}rkner, P.-C.
  (2021).
\newblock Rank-normalization, folding, and localization: an improved {R}-hat
  for assessing convergence of {MCMC} (with discussion).
\newblock {\em Bayesian Analysis}, 16(2):667--718.

\bibitem[Wahba, 1984]{wahba1984partial}
Wahba, G. (1984).
\newblock Partial spline models for the semi-parametric estimation of several
  variables.
\newblock In {\em Statistical Analysis of Time Series, Proceedings of the Japan
  US Joint Seminar}, pages 319--329.

\bibitem[Wand, 2000]{wand2000comparison}
Wand, M.~P. (2000).
\newblock A comparison of regression spline smoothing procedures.
\newblock {\em Computational Statistics}, 15:443--462.

\bibitem[Wood, 2003]{wood2003thin}
Wood, S.~N. (2003).
\newblock Thin plate regression splines.
\newblock {\em Journal of the Royal Statistical Society: Series B (Statistical
  Methodology)}, 65(1):95--114.

\bibitem[Wood, 2016]{wood2016just}
Wood, S.~N. (2016).
\newblock Just another {Gibbs} additive modeller: interfacing {JAGS} and mgcv.
\newblock {\em Journal of Statistical Software}, 75(7):1--15.

\bibitem[Zimmerman and Ver~Hoef, 2021]{zimmerman2021deconfounding}
Zimmerman, D.~L. and Ver~Hoef, J.~M. (2021).
\newblock On deconfounding spatial confounding in linear models.
\newblock {\em The American Statistician}, 76(2):159--167.

\end{thebibliography}
	
\end{document}